    \documentclass[]{aa}
    \usepackage{graphicx}
    \usepackage{natbib}
    \usepackage{longtable}

    \def\Ni{\noindent}
    
    \def\ergsec{\hbox{erg s$^{-1}$ }}
    \def\ergcm{\hbox{erg cm$^{-2}$ s$^{-1}$ }}
    \def\erga{\hbox{erg cm$^{-2}$ s$^{-1}$ \AA$^{-1}$ }}
    
    \def\msun{$M_{\odot}$}
    \def\Msun{M$_{\odot}$}
    \def\rsun{$R_{\odot}$}

    \def\phiorb{\ifmmode\phi_{\rm orb}\else$\phi_{\rm orb}$\fi}
    
    \def\it{\sl}
    \def\degs{\ifmmode ^{\circ}\else$^{\circ}$\fi}
    \def\amin{\ifmmode ^{\prime}\else$^{\prime}$\fi}
    \def\asec{\ifmmode ^{\prime\prime}\else$^{\prime\prime}$\fi}

    \def\ob{\mbox{HU Aqr}}

\begin{document}

     \title{Hunting high and low: XMM monitoring of the eclipsing polar
     \mbox{HU Aquarii}
	\thanks{Based on observations obtained with XMM-Newton, an ESA
     science mission with instruments and contributions directly funded 
     by ESA member states and NASA.}
     }

       \author{R. Schwarz\inst{1} 
               \and
               A.~D. Schwope\inst{1} 
               \and
               J. Vogel\inst{1} 
               \and
              V.~S. Dhillon\inst{2}
               \and
              T.~R. Marsh\inst{3}
               \and
              C. Copperwheat\inst{3}
               \and
              S.~P. Littlefair\inst{2}
               \and
              G. Kanbach\inst{4}
             }

      \authorrunning{R.~Schwarz et al.}
  \titlerunning{XMM monitoring of the eclipsing polar \ob }
   \offprints{R.Schwarz, rschwarz@aip.de}
 
  \institute{
   Astrophysikalisches Institut
          Potsdam, An der Sternwarte 16, D--14482 Potsdam, Germany
        \and 
Department of Physics \& Astronomy, University of Sheffield, Sheffield, S3 7RH, UK 
        \and 
Department of Physics, University of Warwick, Coventry, CV4 7AL, UK
        \and Max-Planck-Institut f\"{u}r Extraterrestrische Physik,
           Giessenbachstra\ss e, D--85740 Garching, Germany
      } 

   \date{Received 2008; accepted 2009}
\abstract{}{We want to study the temporal and spectral behaviour of \ob\ 
in the X-ray domain during different accretion states.}
{We obtained spectra and light curves from four different 
XMM-Newton pointings covering intermediate and low states. The 
X-ray observations were accompanied with high time resolution 
photometry obtained with the Optima and ULTRACAM instruments.}
{On two occasions in May 2002 and 2003 \ob\ was found in an intermediate 
state with the accretion rate reduced by a factor of 50 compared to earlier
high state measurements. X-ray spectra in the intermediate state can be
described by a model containing a blackbody component and hot thermal
plasma. Contrary to the high state the ratio between soft and hard X-ray 
flux is nearly balanced. 
In agreement with previous measurements we observed a migration of the 
accretion spot and stream towards the line connecting both stars.
The brightness of \ob\ was further reduced by a factor of 80 during two low 
states in
October 2003 and May 2005, where it was detected at a luminosity of only
$L_{\rm X} = 4.7\times10^{28}$\ergsec . This luminosity would fit well 
with an active coronal emitter, but the relatively high plasma
temperatures of 3.5 and 2.0 keV are more compatible with residual accretion.
We updated the eclipse ephemeris of \ob\ based on the eclipse egress
of the accretion spot
measured  in various wavelength bands. The $(O-C)$-diagram  of the observed 
accretion spot eclipse timings reveals complex deviations from a linear 
trend, which can be explained by a constant or cyclic period change or 
a combination thereof. 
The quadratic term implies a period decrease at a rate of
$\dot{P}_{\rm orb} = -7..-11 \times 10^{-12}$ s~s$^{-1}$. In case the 
observed period change reflects a true angular momentum loss, this would 
be a factor of 30 larger than given by gravitational radiation. 
}{}
   
\maketitle
\section{Introduction}
Polars or AM Herculis stars are important 
objects to understand accretion in the presence of a strong magnetic field.
In these Cataclysmic Variables \citep{Warner95} the field of the white dwarf 
primary is sufficient to lock its rotation to the binary orbit, and to 
prevent the formation of an accretion disc.
To date, we know 15 eclipsing systems of this kind, which additionally 
provide constraints on the viewing geometry and allow the different 
emission sites to be disentangled.
Among those, HU Aqr is one the of the brightest  and
has become one of the most comprehensively studied systems. 
Since its discovery in the ROSAT survey \citep{Schwope93} it has 
been studied in various wavelength bands,
and was the first system in which the ballistic stream was detected
via Doppler tomography \citep{Schwope97}.
It features a rather simple accretion geometry with one dominating 
accretion spot that undergoes self-eclipses. 
It was the target of a long-term monitoring programme 
with the ROSAT and EUVE satellites
\citep{Schwope01}. Part of these observations were obtained in a high state 
that revealed a strong soft X-ray excess and could be used to 
resolve the eclipse egress, thus directly constraining the diameter of the
soft X-ray emitting area to $\sim 450 $km.  
Modelling of the X-ray eclipse during the high state located the accretion
region 
at azimuth $\psi = 46$\degr\ and colatitude $\beta = 31$\degr\ and required a 
substantial amount of vertical extent.
During the monitoring \ob\ displayed large variations of the accretion rate,
accompanied by a correlated migration of the accretion 
spot and the coupling region. For the reduced (intermediate) state of
accretion both accretion spot and stream were shifted towards the line
connecting both stars in agreement with an increased threading radius 
given by the magnetic/ram pressure balance.
All of the intermediate state observations
where obtained with the ROSAT-HRI and EUVE only, thus lacking spectral
information.
In this paper we present a series of XMM pointings that continue 
the monitoring through intermediate and low brightness states of the system.
Preliminary results of this study were presented in \citet{Schwope04} and
\citet{Vogel08a}.
\section{Observations and reduction}

\begin{table}
\caption{Log of the XMM X-ray observations of HU Aqr}
\begin{tabular}{llcc}
\noalign{\smallskip} \hline \noalign{\smallskip}
      Date & Detector$^{(1)}$  & $T_\mathrm{Exp}$  
      & Mean rate$^{(2)}$  \\
     & &  (ksec)        &    (cts/s)  \\
\noalign{\smallskip} \hline \noalign{\smallskip}
    2002 May 16--17 & PN-sw  & 36.4&  1.11   \\
                    & MOS1-sw  & 37.2 & 0.19       \\
                    & MOS2-sw  & 37.2 & 0.26      \\
    2003 May 19     & PN-ti  & 18.5 &  0.94   \\
                    & MOS1-ti  & 20.2   & 0.21    \\
                    & MOS2-ti  & 20.0   & 0.18     \\
    2003 Oct 24--25 & PN-sw  & 11.4   & 0.011  \\
                    & MOS1-sw  & 11.6   &  -\\
                    & MOS2-sw  & 11.6   &  -\\
    2005 May 16     & PN-ti  & 18.6  & -\\
                    & MOS1-sw  & 19.4   & 0.0025     \\
                    & MOS2-sw  & 19.4   & 0.0022\\
 \noalign{\smallskip}
 \hline
 \noalign{\smallskip}
      \end{tabular}\\
   \noindent{\Ni\small $^{(1)}$ The abbreviations after the PN/MOS detector 
   indicate the instrument mode: sw = small window, ti = timing\\ 
                       $^{(2)}$ In the bright phase interval $\phi =
                       0.62-0.91$ and $\phi = 0.05-0.18$. Energy range used 
                       0.18--10 keV and 0.3--10 keV for imaging and timing 
                       modes, respectively. Count rates have been corrected
                       for the 65\% encircled energy fraction corresponding 
                       to the aperture used.
            }
    \label{xlog}
   \end{table}

Between 2002 and 2005 \ob\  was the target of four dedicated XMM pointings.
Different instrument configurations of the EPIC
cameras (Table~\ref{xlog}) were used to best match the brightness 
level of the source. 

For the MOS and PN data taken in image mode source photons have been 
extracted within a circular aperture of 15\arcsec . The resulting 
count rates given in the paper have been corrected for the 65\% encircled
energy fraction corresponding to this radius. The background level has been 
determined from a larger source-free region close to the source. 
For the timing mode
observation the level of background radiation was comparable to the 
source flux resulting in much poorer data quality. 
The PN-data additionally suffered from soft proton flares, so here all 
photons with energies lower than 0.3 keV had to be omitted.
Apart from this particular data set, spectral analysis was applied to 
the energy range between 
0.18--10 keV including good events (flag = 0) identified with patterns 
0--12 and 0--4 for the MOS and PN-CCD, respectively. We used version 
7.0.0 of the XMMSAS package.

During each pointing the Optical Monitor (OM) was operated with the filter UVM2 
using the fast mode. These data were reduced using the task {\sc omfchain}.
For conversion of the count rates to flux units we applied the factors 
given in the XMM calibration file OM\_COLORTRANS.

We also obtained contemporary optical data with the fast-timing photometric 
devices ULTRACAM \citep{Dhillon07} and Optima \citep{Kanbach03}.
These instruments were operated at 
the 4.2-m William Herschel Telescope (WHT) on La Palma, the VLT on Paranal, 
and the 1.3 m telescope of the Skinakas Observatory, Crete.
A detailed description of these optical and the OM UV observations will be 
given in a different paper (Vogel et al., in prep.).

\section{Results}
\subsection{An updated accretion spot ephemeris}
\label{s:ephe}

\begin{figure}
\includegraphics[width=1.0\columnwidth,clip]{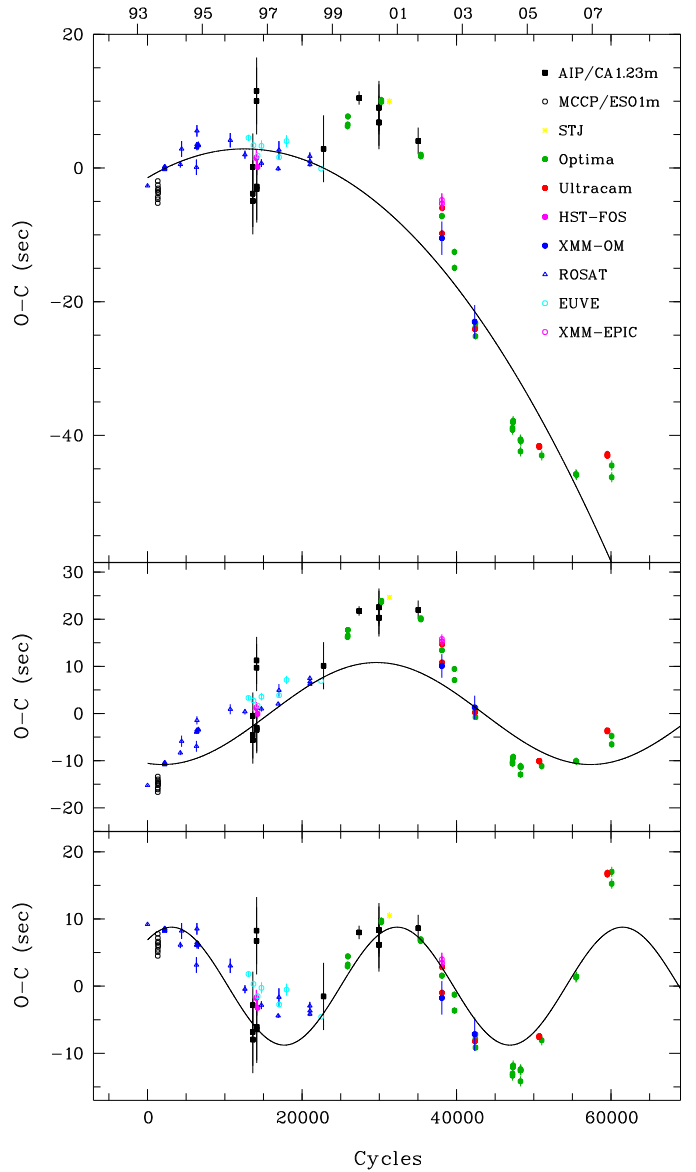}

\caption{Upper panel: Residuals of the timings of accretion spot egress with 
respect to the linear ephemeris of \cite{Schwope01} together with the best-fit 
quadratic ephemeris derived in this paper (solid line).
Middle panel: Residuals with respect to  
our new linear ephemeris together 
with the best-fit linear+sinusoidal ephemeris (solid line).
Lower panel: Residuals with respect to the linear and quadratic terms of the 
combined quadratic and sinusoidal ephemeris. Overplotted as a solid line 
is the sinusoidal term of this ephemeris.}
\label{f:omc}
\end{figure}
We have updated the eclipse ephemeris of HU Aqr using 
newly available timings of the spot egress measured in 
the optical, UV and X-rays (Table~\ref{t:timings}) and 
the  older ROSAT and EUVE measurements of \citet{Schwope01}. 
As the timestamp of egress/ingress in the  optical 
we used the moments of half intensity. 
For the photon counting data (UV and X-ray) this datum was 
computed from the mean of the arrival times of the first three photons
after the eclipse.
Excluded were the low state OM light curves, which are of low 
counting statistic and likely dominated by the white dwarf photosphere.
All timings have been converted to the solar system's barycenter and are 
in the time frame of ephemeris time which accounts for leap seconds.

\begin{table}
  \caption{Eclipse egress ephemerides of HU Aqr}
  \begin{tabular}{ll}
    \hline
    \multicolumn{2}{l}{\bf Linear ephemeris:} \\
    \multicolumn{2}{l}{BJED = T$_{0}$ + P$_{0}\cdot E$} \\ [1ex]

    T$_{0} = 2\,449\,9102.9200839\,(\pm 6) $& P$_{0} = 0.08682041087\,(\pm 3)$ d\\
    $\chi^{2}_{\nu}= 345.9,
    \;\;\; \nu$ = 101 & \\ [2ex] 
    
    \multicolumn{2}{l}{\bf Quadratic ephemeris:} \\
    \multicolumn{2}{l}{BJED = T$_{0}$ + P$_{0}\cdot E$ + $Q\cdot E^{2}$} \\ [1ex]
    T$_{0} = 2\,449\,102.9200164\,(\pm 8)$ & P$_{0} = 0.08682042416\,(\pm 8) $ d \\
    Q $ = (-3.17 \pm 0.02) \times 10^{-13}$  &  \\
    $\chi^{2}_{\nu}= 40.5,
    \;\;\; \nu$ = 100 & \\ [2ex]
    
    \multicolumn{2}{l}{\bf Sinusoidal ephemeris:} \\
    \multicolumn{2}{l}{BJED = T$_{0}$ +
      P$_{0}\cdot E$ + A$\cdot \sin\,[2\pi (E-{\rm B})/{\rm C}]$} \\ [1ex]
T$_{0}=  2\,449\,102.9201788\,(\pm 15)$ d & P$_{0}= 0.08682040612\,(\pm 6)$ d \\
A $= (125 \pm 1) \times 10^{-6}$ d & B $= (89 \pm 4) \times 10^{3}$ cycles \\
C $= (554 \pm 5) \times 10^{2}$ cycles  & \\ 
    $\chi^{2}_{\nu}= 30.4,
    \;\;\; \nu= 98$ & \\ [2ex]

    \multicolumn{2}{l}{\bf Quadratic + sinusoidal ephemeris:} \\
    \multicolumn{2}{l}{BJED = T$_{0}$ + P$_{0}\cdot E$ + $Q\cdot E^{2}$ + 
      A$\cdot \sin\,[2\pi (E-{\rm B})/{\rm C}]$} \\ [1ex]
    T$_{0} = 2\,449\,102.9198961\,(\pm 25)$ & P$_{0} = 0.0868204353\,(\pm 2) $ d \\
    Q $ = (-4.77 \pm 0.04) \times 10^{-13}$  & A $= (101 \pm 2) \times 10^{-6}$ d \\ 
B $= (-4190 \pm 150) $ cycles & C $= (291 \pm 1.5) \times 10^{2}$ cycles \\  
    $\chi^{2}_{\nu}= 11,
    \;\;\; \nu$ = 97 & \\ [2ex]
    \hline

  \end{tabular}
  \label{t:allephem}
\end{table}

In Fig.~\ref{f:omc} we compare the observed egress timings with the 
predictions of the linear ephemeris of \cite{Schwope01}, which reveals
conspicuous deviations of up to 45 sec for the latest observations. 
We have updated the linear ephemeris using a least-squares fit weighted by 
the inverse of the squares of the eclipse timings. The best-fit parameters 
together with their formal 1-$\sigma$ errors are given in 
Table \ref{t:allephem}.
The fit is very poor, showing large positive and negative residuals.

A better match to the data can be achieved by introducing either a quadratic 
or a sinusoidal term. The first would indicate a secular variation of the 
orbital period, while the latter indicates a cyclical period change. 
The quadratic term is large and would imply an orbital period 
decrease of $\dot{P}_{\rm orb} = -7.3 \times 10^{-12}$s~s$^-1$, which is
larger than what is expected from angular momentum 
loss due to gravitational radiation. The cycle length of the sinusoidal 
modulation is 13 yr, close to the time-span of our observations. 
The quadratic and sinusoidal ephemerides are shown as solid lines in the 
upper and middel panels of Fig.~\ref{f:omc}.
Both models feature a large $\chi_{\nu}$ of 40.5 and 30.4 emphasizing that
the period variation is more complex than these simple fits. 
Of particular note is a rapid increase in
the $(O-C)$-values of 10 sec/yr between 1998 and 2000, 
and a subsequent strong period decrease in the following years lasting until
2004. This could be indicative of a cyclical period modulation on a shorter
time-scale of just a few years.
As a possible model we assumed a combination of a quadratic plus 
sinusoidal term, which account for a long-term period change 
and cyclical modulation simultaneously. 
The result of this fit is displayed in the lower panel of Fig.~\ref{f:omc}, 
where the residuals of the linear and quadratic part of the ephemeris 
are shown together with the fitted sinusoidal term. 
The combined fit yields a $\dot{P}_{\rm orb}$ of $-1.1 \times
10^{-11}$s~s$^-1$,
slightly larger compared to the simple quadratic ephemeris. 
The cyclical variation has a period of 6.9 yr and an semi-amplitude 
of $\sim 9$ sec.
Our data so far faithfully revealed two consecutive cycles of this putative 
modulation lending some support for its existence. On the other hand, the very 
latest data points are already deviating by 10 sec suggesting that 
the derived period decrease might be to large.  

With a $\chi_{\nu}$ of 11.5 the fit is clearly better than the
single quadratic or sinusoidal models. Using an $F$-test following 
the prescription of \cite{Pringle75} the significance
of the quadratic plus sinusoidal ephemeris compared to the linear plus 
sinusoidal ephemeris is $93.9$ per cent with $F(1,97) = 170$.
While this model is statistically favoured, it is too early to say 
what the future behavior of the period wanderings in \ob\ will be. 
However, 
the eclipse timings to be taken in the next few years should clearly 
discriminate between the different possible solutions and confirm the
existence of a large quadratic term.

\begin{figure}
\includegraphics[width=1.0\columnwidth,clip]{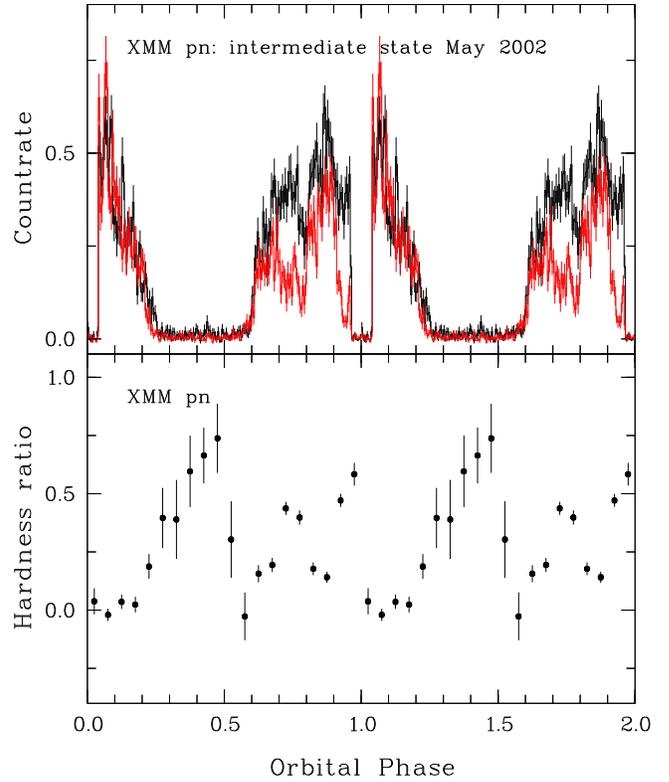}
\caption{
upper panel: Phase-averaged EPIC PN
X-ray light curve of \ob\ observed with XMM-Newton in May 2002. 
The black and gray line mark the light curve in the hard (0.5--10 keV) and 
soft (0.18--0.35 keV) energy bands. lower panel:
Variation  of the hardness ratio ((H-S)/(H+S)) as a function of orbital phase.
}\label{f:xlc02}
\end{figure}
\begin{figure}
\centerline{\includegraphics[width=0.90\columnwidth,clip,angle=-90]{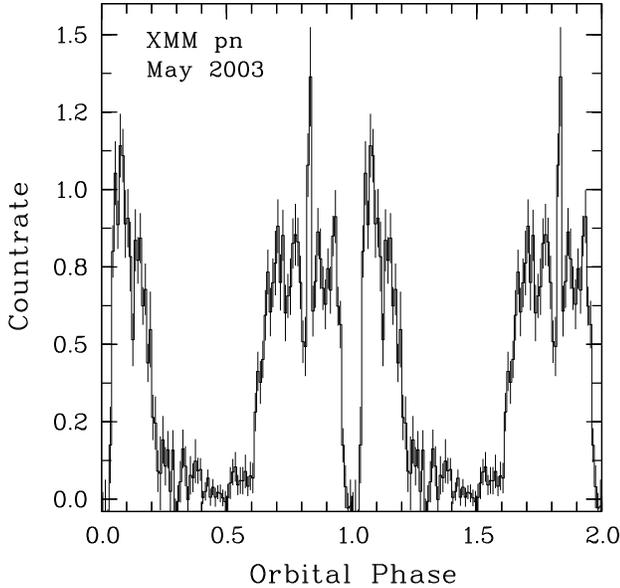}}
\caption{
Phase-averaged EPIC PN X-ray light curve of \ob\ in the 0.3--10 keV 
range observed with XMM-Newton in May 2003. 
}\label{f:xlc03}
\end{figure}

\begin{figure}
\includegraphics[width=0.75\columnwidth,angle=-90,clip]{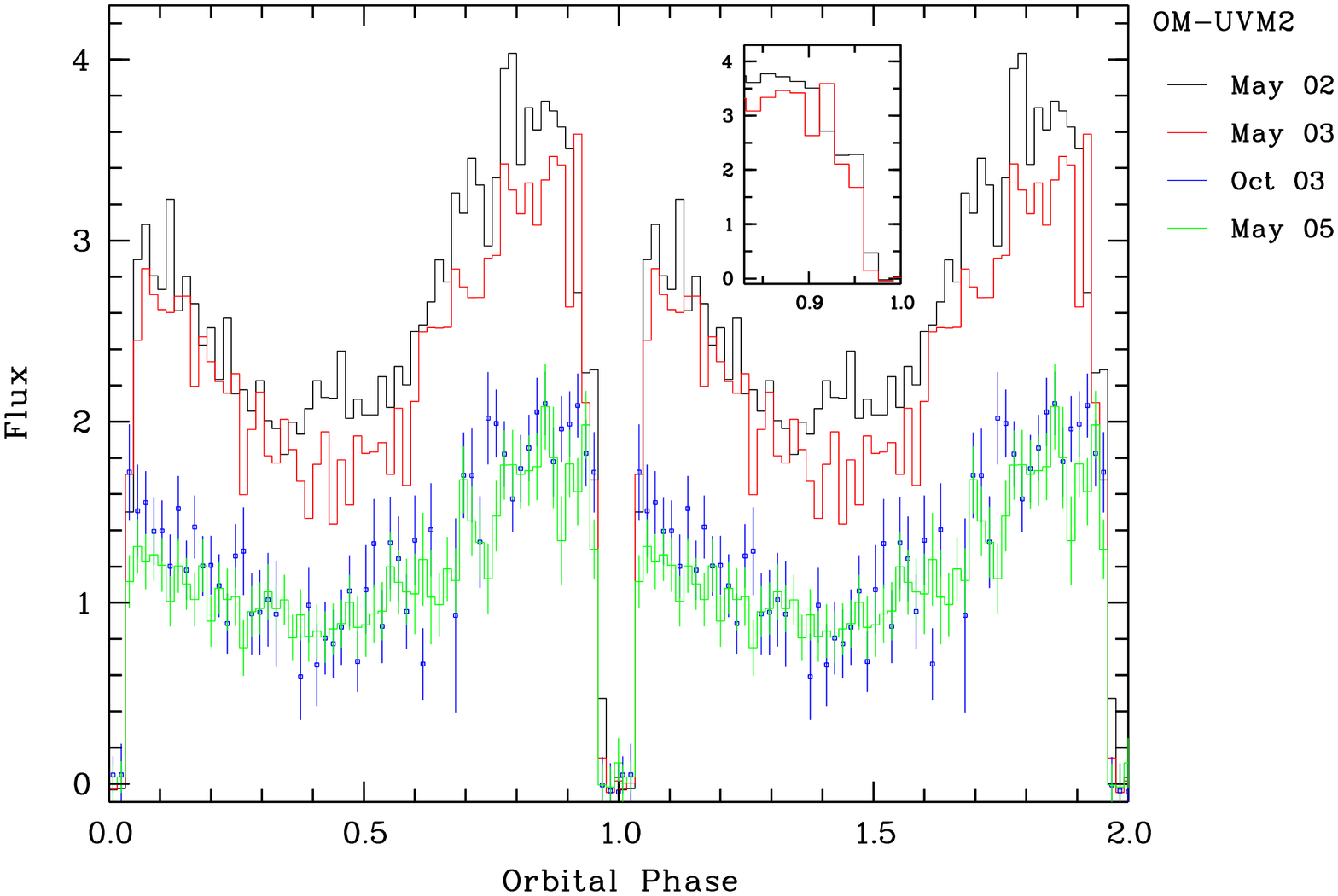}
\caption{ 
Phase-averaged OM light curves observed in the UVM2 filter at 2200 \AA . 
The two upper curves are from 
the intermediate states of May 2002 (black) and May 2003 (grey), while 
the two lower graphs correspond to the low states of October 2003 (dots) and
May 2005 (histogram). The flux unit is $10^{-15}$\erga . 
}\label{f:omlc}
\end{figure}
 
\subsection{X-ray and UV light curve morphology}
In Fig.~\ref{f:xlc02} and Fig.~\ref{f:xlc03} we show the EPIC-PN light
curves during the intermediate states observed in May 2002 and May 2003.
For the data from May 2002 we extracted photons in two spectral bands 
ranging from 0.18--0.35 and 0.5--10 keV, which match the two spectral 
components (see below). 
A corresponding hardness ratio 
has been calculated with the two bands related as (hard-soft)/(hard+soft). 

As in previous X-ray observations the light curves are dominated
by a single bright phase interval of a  self-eclipsing pole.
The centre of the bright phase of the main 
accretion region was at $\phi =0.912$ and $\phi =0.915$ for the May 2002
and 2003 observation, respectively.
This corresponds to an azimuthal position of $\psi \simeq 31\degr$.
There is no obvious difference between the location of the soft and 
hard X-ray emitting region.
Compared to previous observations these new measurements show the main 
accretion region 
most closely shifted towards the line connecting both stars. 
Such a relation between position and 
instantaneous accretion rate is expected for an intermediate state 
from the general trend seen in some older ROSAT/EUVE observations 
\citep{Schwope01}.

The light curve from May 2002 reveals the  
presence of narrow and broad absorption dips in the soft X-ray 
light curve.  
For the observation of May 2003 
the counting statistic in the soft band is not sufficient for
a clear detection of either dip.
In May 2002 the narrow absorption feature was centred 
at phase $\phi=0.941$ with the dip ingress starting at $\phi=0.908$. 
These values are among the closest to the eclipse observed, 
indicating a migration 
of the stagnation point towards the line connecting both stars.
A broad absorption trough is centred around $\phi = 0.73$ very 
similar to what was previously observed in several ROSAT and EUVE 
pointings.

There is a substantial
hardness ratio variation during both the narrow and the broad absorption
dip. The narrow dip is not total even in the softest X-ray band. 
A fit to the dip  spectrum with the temperature of the 
blackbody components fixed yields a cold absorption column 
of $(2.7 \pm 0.2)\times10^{20}$cm$^{-2}$. This is a factor of ten 
lower then the $(2-3)\times10^{21}$cm$^{-2}$ measured during the 
ROSAT high-state observation from October 1993.

Contrary to the bright-state ROSAT observation from October 1993, 
eclipse ingress and egress profiles are unresolved for the XMM observations
at a timescale shorter than $\sim 2$ sec due to the low count rate 
of the data.

Fig.~\ref{f:omlc}  shows the phase folded light curves of \ob\
taken with the Optical Monitor and the UVM2 filter. The source was 
in two different brightness states, being brighter by a factor
of two during the intermediate state observed in May 2002 and
May 2003. During both low states \ob\ was at the same brightness level,
between $0.8\times10^{-15}\erga$ in the faint phase 
and $2\times10^{-15}\erga$ at orbital maximum.  
At all four epochs we observe a smooth sine-like modulation 
reminiscent of a heated cap on the surface of the white dwarf
\citep{Gaensicke06}.
The level of modulation 
is rather similar for all brightness states. 
This suggests that the amount of heating does not strongly depend on the 
instantaneous irradiation.
A detailed modelling of this feature will be presented in 
Vogel et al., in prep.

In the intermediate brightness state the pre-eclipse UV flux 
is depressed  by 30\% starting at phase 0.92 
(see inset of Fig.~\ref{f:omlc}).
A plausible explanation for this dip is absorption by the accretion stream 
also seen in soft X-rays.
The additional brightness compared to the low state is 
seen in the bright and faint phase, i.e. it does not come from 
the self-eclipsing heated cap. 
A possible source of this emission is the accretion stream, 
which is also indicated by a prolonged ingress after eclipse 
of the white dwarf.

\subsection{X-ray spectroscopy}
\label{s:xspec}

\begin{figure}[tb]
\begin{center}
\includegraphics[clip=,width=0.98\columnwidth]{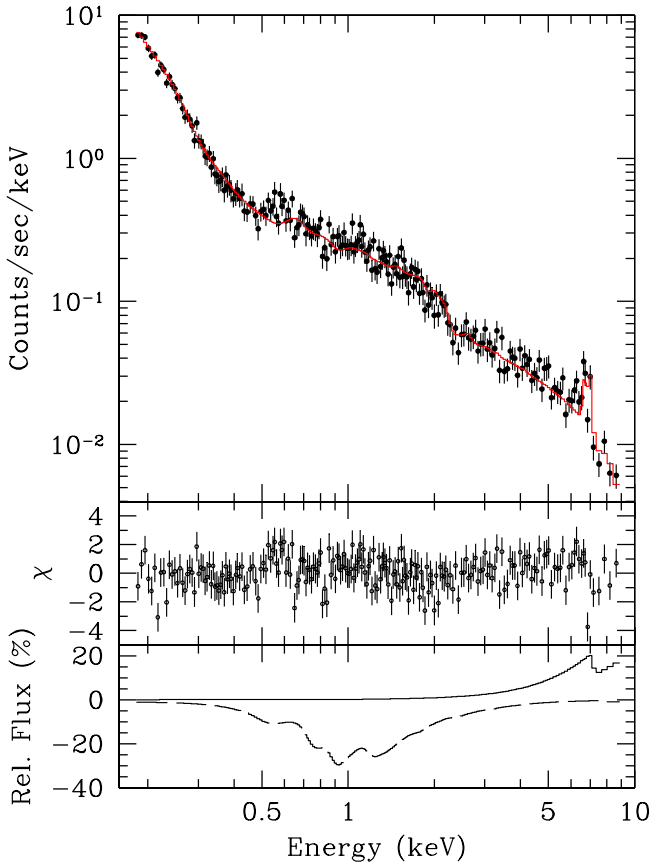}
\caption{
EPIC-PN bright phase spectrum of \ob\ taken in May 2002 together with a fit 
containing a blackbody and mekal model modified by ionised absorption and 
cold reflection. {\it Center:} Residuals with respect to the fit. 
{\it Bottom:} Relative contribution of the ionised absorber (dashed line) and 
reflection component (solid line) to the total X-ray spectrum.}
\label{f:xspec}
\end{center}
\end{figure}
X-ray spectra in the intermediate state have been extracted using the bright 
phase intervals from $\phi = 0.62-0.91$ and $\phi = 0.05-0.18$,
thus excluding the absorption dip and the eclipse.  
The spectra of the PN and MOS detectors have been fitted together and 
can be described with a two component model including  
a soft blackbody and hot thermal plasma.
For the latter 
both a {\sc mekal} 
or a bremstrahlung model fits relatively well with a $\chi^{2}$ of 1.46 and
1.49.  
The observation from May 2002 is sufficient to resolve individual line
features.
Above 6 keV all three lines of fluorescent, H-like and He-like iron are 
present and do contribute to the $\chi^{2}$. 
We  also note  weak residuals seen at 0.5 keV from a
blend of line emission, probably dominated by the \ion{O}{vii} and 
\ion{O}{viii} lines at 21.8 and 19 \AA .  
The ratio between the
H-like and He-like iron lines
indicates plasma with temperatures between 5--15 keV, while the 
continuum slope at energies larger 5 keV forces the fit to unrealistically
high temperatures
of $kT_{\rm brems} = 115$ keV and $kT_{\rm mek} =98$ keV.
Introducing a continuous temperature {\sc cmekal} model does not help to 
solve the discrepancy between the temperature indicators. 

A way to reconcile 
the spectral slope with more realistic plasma temperature  
is by adopting an ionised absorber as implemented in {\sc XSPEC} by 
the {\sc absori} model \citep{Done92}.
Introduction of the {\sc absori} model significantly improves the fit to
$\chi_{\nu}^{2}=1.38$. Using an $F$-test this model is better than 
the model without warm absorption at the 75\% significance level. 
Comparison with fits to 
the PN-data alone yields a $\chi_{\nu}^{2}$ of 1.12, showing that the 
remaining discrepancy is mainly due to the uncertainty in the 
cross-calibration of the different detectors.
As parameters of the {\sc absori} model  we fixed the power law index 
of the incident radiation to 1.25 and the temperature of the ionised 
material to 30000 K.  The remaining best fit parameters then are 
an  absorption column of $6.5\times 10^{21}$cm$^{-2}$ 
and an ionisation parameter $\xi = 110$ erg~cm~sec$^{-1}$.
For the observation of May 2003 these parameters were fixed at the 
values from 2002 given the low quality of the data.
The complex absorption drastically decreases the best-fit temperatures 
of the {\sc mekal} model to 19 and 15 keV for 
the observation in 2002 and 2003, respectively.

The incidence of the fluorescent iron line at 6.4 keV reveals that
there is also reflection from neutral material at the surface of the white 
dwarf. This has been taken into account by adding 
the {\sc reflect} model \citep{Magdziarz95} to our fitting.
The new model component only marginally improves the $\chi_{\nu}^{2}$, 
but lowers the  temperatures of the {\sc mekal} component to 15 and 13 keV
for the observation in 2002 and 2003, respectively.
The only free paramter is the scaling factor of the 
reflected component, which for the fit of the 2002 data is  $R=1.66$.
This value was fixed 
for the fit to the much poorer data from 2003. 
For completeness we also provide a fit using only the reflection model   
without an ionised absorber. For the 2002 observation this fit has a 
significantly higher $\chi_{\nu}^{2}$ compared to the fit including warm 
absorption, while the difference between both models is negligible 
for the data of May 2003. 
Using this model the best-fit temperatures of the {\sc mekal} component 
change to 35 and 22 keV for the observation in 2002 and 2003, respectively.

The blackbody temperatures
for the observations of May 2002 and May 2003 differ and also depend on 
the details of the absorption and reflection models used.  
The best-fit values for our finally adopted model including warm absorber 
and reflection  are $kT_{\rm bb} = 34$ eV and $kT_{\rm bb} = 25$ eV, 
differing only slightly above the $2\sigma$-level. 
Those values fall right into the range of temperatures 
measured for polars with a dominant soft component \citep{Ramsay03,Ramsay04c}.
The decrease  of the temperature in May 2003 is accompanied by an 
increase of the unabsorbed, bolometric soft X-ray flux by a factor of 6. 
At the same time the bolometric hard X-ray flux decreased by 10\% . 
A much more significant change (at the $3\sigma$ level) 
of the blackbody temperature is observed if one compares
the 34 eV of the intermediate state in 2002 with the 22 eV measured
during the previous high state in October 1993 \citep{Schwope01}.

We also note the change of the hydrogen column density compared to the
high state ROSAT observation. Both intermediate state observations require 
at maximum a density of 
$6\times10^{19}$cm$^{-2}$, whereas the ROSAT pointing is consistent with a 
larger column of $2\times10^{20}$cm$^{-2}$. The large variability of the
absorption column point toward an intra-binary origin.

Using the values from the fit including absorption and reflection,
and taking $\pi$ and $2\pi$ for the scaling of the blackbody and 
thermal emission, the unabsorbed flux ratio 
$f_{\rm bol,soft}/f_{\rm bol,hard}$ were 1.3 and 8.4
for the May 2002 and May 2003 intermediate states. Taking into account  
the uncertainty due to the unknown emission characteristic and the neglect of 
the cyclotron component, the value from May 2002 is relatively close to the 
ratio of $\sim 0.5$ of the standard accretion shock model of \citet{Lamb79}
and \citet{King79}. Both values mark a dramatic decrease of the soft X-ray 
excess compared to the high state of October 1993, where the soft-to-hard ratio
of the bolometric fluxes was $\sim 800$. Note that this change is almost 
solely due to the decrease of 
the soft X-ray component, whereas the hard X-ray flux remained nearly 
constant from the high to the intermediate state. 

The total luminosity in X-rays $L_{\rm X} = (\pi f_{\rm bb}+2\pi f_{\rm
therm}) D^{2}$, assuming a distance of 180 pc, was $2.5\times 10^{31}$ and 
$9.3\times 10^{31}$ \ergsec\
in May 2002 and May 2003. The inferred accretion rate is between 
$(0.8-3.1)\times
10^{-11}$\Msun /yr , a factor of 20 lower compared to the high state.  

   \begin{table*}
     \caption{Fit results for the combined PN and MOS X-ray spectra of \ob}
      \begin{tabular}{llcccccl}
      \hline
      \noalign{\smallskip}
       Model & $N_{\rm H}^1 $ 
                  & $kT_{\rm bb}$ 
                  & $kT_{\rm therm}$ 
                & $F_{\rm soft,bol}$
                & $F_{\rm hard,bol}$
                & $\chi^2_{red}$ \\
          &  (10$^{19}$ cm$^{-2}$)  & 
         (eV) &
         (keV) &
         $({\rm erg~cm}^{-2}{\rm s}^{-1})$  & 
         $({\rm erg~cm}^{-2}{\rm s}^{-1})$  & \\
       \noalign{\smallskip}
      \hline
      \noalign{\smallskip}
{\it May 2002}   & & & &\\
 bb+brems    & $0^{+2}_{-0}$   & 35$^{+3}_{-3}$&115$^{+15}_{-15}$&$8.0\times10^{-12}$&$1.4\times10^{-11}$ & 1.49\\
 bb+mek      & $0^{+2}_{-0}$   & 35$^{+1}_{-1}$&98$^{+10}_{-10}$&$8.0\times10^{-12}$ &$1.4\times10^{-11}$ & 1.46\\
 reflect(bb+mek)   & $0^{+2}_{-0}$   & 35$^{+1}_{-1}$&$30^{+25}_{-10}$&$8.3\times10^{-12}$ &$7.2\times10^{-12}$ & 1.44\\
absori(bb+mek)   & $3^{+7}_{-3}$  &
 32$^{+2}_{-2}$&$18^{+14}_{-4}$& $1.4\times10^{-11}$ & $6.7\times10^{-12}$& 1.37\\[3pt]
reflect $\ast$ absori(bb+mek)   & $0^{+7}_{-0}$  &
 34$^{+1}_{-3}$&$15^{+6}_{-3}$& $1.5\times10^{-11}$&$5.6\times10^{-12}$ & 1.36\\[3pt]
{\it May 2003}   & & & &\\
bb+brems    & $1.8^{+4}_{-2}$ & 29$^{+4}_{-3}$&69$^{+40}_{-30}$&$2.5\times10^{-11}$& $1.1\times10^{-11}$ & 1.46\\
bb+mek    & $2.0^{+4}_{-2}$   & 29$^{+4}_{-3}$&54$^{+40}_{-30}$&$2.7\times10^{-11}$& $1.0\times10^{-11}$& 1.43\\
reflect(bb+mek)   & $2.7^{+4}_{-2}$  & 29$^{+4}_{-3}$&$22^{+10}_{-6}$&
 $3.4\times10^{-11}$ &$5.9\times10^{-12}$ & 1.38\\
absori(bb+mek)   & $5.9^{+3}_{-3}$  &
 25$^{+3}_{-3}$&$15^{+5}_{-3}$& $8.7\times10^{-11}$ & $5.7\times10^{-12}$& 1.39\\[3pt]
reflect $\ast$ absori(bb+mek)   & $5.8^{+6}_{-3}$  &
 25$^{+3}_{-3}$&$13^{+3}_{-2}$& $8.6\times10^{-11}$ & $5.1\times10^{-12}$& 1.36\\[3pt]
{\it Oct 2003}   & & & &\\
mek    & $2^{+10}_{-2}$   & &$3.5^{+3}_{-3}$&& $4.8\times10^{-14}$& 0.5\\[3pt]
{\it May 2005}   & & & &\\
mek    & $0^{+10}_{-0}$   & &$ 1.96^{+1}_{-1} $&& $4.3\times10^{-14}$& 1.25\\[3pt]
\noalign{\smallskip} \hline \noalign{\smallskip}
\end{tabular}

\noindent{\small 
$^1$ The $N_{\rm H}$ refers to cold absorption applied to all model
components. 
}
   \label{fitres}
   \end{table*}
\subsection{Deep low-state observations}

\begin{figure}[tb]
\begin{center}
\includegraphics[clip=,width=0.98\columnwidth]{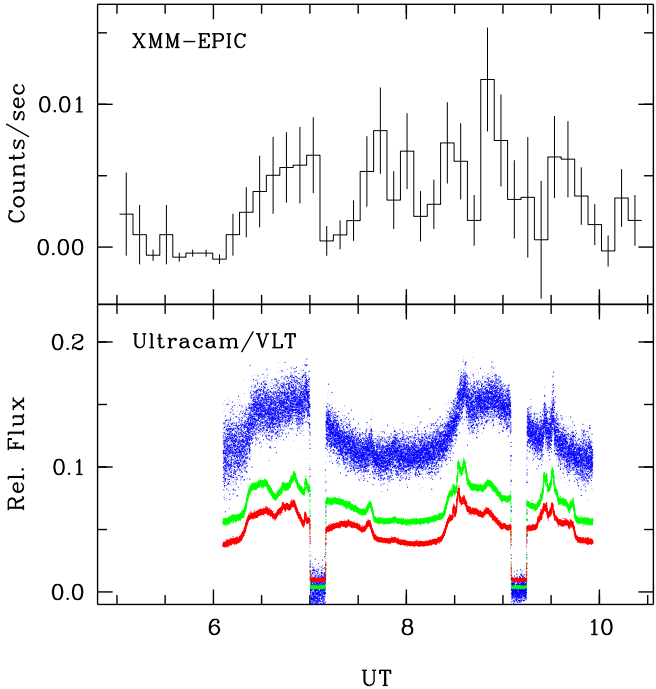}
\caption{
Light curves of the simultaneous XMM-MOS/ULTRACAM observation of \ob\ on
May 16, 2005. The relative flux units of the
optical light curves are with respect to the comparison star of
\citet{Schwope93}. The blue, green and red lines denote 
the light curves in the $u^{\prime}$, $g^{\prime}$ and $r^{\prime}$ filters.
}
\label{f:lowlc}
\end{center}
\end{figure}
In October 2003 \ob\ was found in a deep X-ray low-state. 
An optical light curve taken on October 21, 2003 three days in advance 
of the X-ray observations showed \ob\ with a bright phase magnitude
of only $R = 17\fm4 $ .
The source was detected with the PN detector 
at a rate of only $0.011\pm 0.003$ cts/sec. 
The spectrum of the $\sim 90$ source photons
can be fitted with a thermal plasma {\sc mekal} model with a 
$\chi_{\nu}^2$ of  0.5. A fit using Cash statistics led to similar results. 
No dedicated soft blackbody component is
required for the low state of \ob . The fit temperature of $3.5\pm   3$ keV 
is not very well constrained.
The unabsorbed bolometric flux of the observed spectrum is 
$4.8\times10^{-14}$\ergcm\ yielding  an X-ray 
luminosity of $4.6\times 10^{28}$\ergsec\ for a distance of 180~pc. 
There is weak indication for a variability pattern, where the photons
cluster in the X-ray bright phase. 

A second deep low state was observed in May 2005.
This observations was accompanied by a series of ULTRACAM light curves 
taken on consecutive nights between May 14 and 17, 2005. 
In the $g^{\prime}$ and 
$r^{\prime}$-band residual accretion activity was evident by cyclotron beaming
and flaring. The $u^{\prime}$-band, on the otherhand, was clearly dominated by
sine-like variations reminiscent of a heated white dwarf cap.
Starting from a  bright phase magnitude of $r^{\prime} = 17\fm5 $ 
seen on May 14, 2005 the level of additional accretion activity increased 
from night to night. 
For the simultaneous observations on May 16, 2005 the maximum brightness was at 
$r^{\prime} = 17\fm3 $. \ob\ was detected in both MOS detectors at mean rates of 
0.0022 and 0.0025 cts/sec.
The X-ray light curve (Fig.~\ref{f:lowlc}) reveals clear variability, which was 
only partially correlated with the visibility of the main accretion region. 
For instance the source appears X-ray bright during one faint phase interval. 
As for the first low state observation the spectrum 
can be sufficiently fitted  by a single temperature thermal plasma model.
The fit parameters imply a negligible amount of absorption and a 
temperature of $kT_{\rm mek} = 2\pm1$ keV. The source was at a similar 
brightness level compared to the October 2003 low state with a 
bolometric flux of $4.3\times10^{-14}$\ergcm .

The observed X-ray emission can be compared with that of single, active
late type stars \citep{Pizzolato03}. Assuming a bolometric luminosity
of $3.9\times 10^{31}$\ergsec\ for the M4.5 secondary 
\citep{Leggett96} results in $L_{\rm X}/L_{\rm bol}=1.2\times 10^{-3}$. 
This value is compatible
with the saturated X-ray regime observed for the fastest rotators, although 
it falls into the top end being a factor of two higher than the mean.
An argument against a coronal origin are the relatively high plasma 
temperatures, that lie above the peak of $log(T({\rm K})) = 7$  in 
the temperature distribution  of coronal emitters \citep{Sanz-Forcada03} and 
are also higher than the temperatures of $\sim 0.6$keV observed in 
low accretion rate polars 
\citep{Vogel07}. The plasma temperatures measured in the low-state of \ob\
are better matched with an 
origin in a low density accretion column that predominantly cools via
cyclotron radiation \citep{Woelk92}. Signs of residual accretion are
obviously present in the $r^{\prime}$ and $g^{\prime}$ light curves 
in May 2005, where cyclotron beaming and flaring is observed.  
The problem with this interpretation is that one would expect a clear 
on/off variability pattern in X-rays due to the self-eclipse of the accretion
region, which is not observed at least for the low state in May 2005. 

\section{Discussion}
\label{s:discuss}
One of the major results of our study is the dependence of the soft X-ray 
excess on the X-ray brightness level and thus the instantaneous accretion 
rate. Unlike the high state when a strong soft X-ray excess was observed,
the system reverted to a more balanced flux ratio in its intermediate 
accretion state. 
A similar result has been reported for a handful of systems in the XMM 
snap-shot survey of polars \citep{Ramsay04}. It can be understood in 
terms of a density change in the coupling region, and is therefore 
an indirect confirmation of 
the blobby accretion scenario \citep{Frank88}.  
Our observation indicates a temperature increase of the blackbody
component from 22 eV to 34 eV as the system changed from the high to
the intermediate state. There are only few observations of polars 
which constrain the physical parameters that determine 
the temperature of the reprocessed X-ray emission. 
One particular example is the ROSAT PSPC observation of AM Her 
\citep{Beuermann08}, which shows a correlated increase of the 
blackbody temperature with increasing soft X-ray flux.  
This result is an indication that the temperature is set in 
response to a locally enhanced accretion rate, at least for this 
given system at certain epoch. 
The anti-correlation of temperature 
and soft X-ray flux in \ob\ would contradict 
this picture and could be explained by a much smaller soft X-ray 
emitting area in the intermediate state. 

During the intermediate state the main accretion spot shifted toward the line
connecting both stars to an azimuth of 
$\psi = 31$\degr .
It thereby follows the accretion-rate dependent migration previously seen in 
some ROSAT/EUVE pointings \citep{Schwope01}. We did not observe the irregular 
behaviour seen in a few ROSAT/EUVE intermediate state observations, where 
the accretion
spot flipped back to the high state position. The accretion stream was detected
via the absorption dip. Compared to the high state its column density was
reduced by a factor  of 10, while its position moved by $\sim 25\degr$ with
respect to its high state position, also in accord with previous observations.

The origin of X-rays during the low state of \ob\ remains ambiguous. While 
the luminosity is consistent with a coronal emitter, the observed 
temperatures are better explained by an accretion plasma. For the latter 
interpretation we would expect a clear orbital modulation which is not
observed.

The $(O-C)$ diagram of the accretion spot eclipse timings in \ob\ reveals
large deviations from a constant period on relatively short timescales. 
It should be noted that the observed spot timings, beside a true change 
of the orbital period, are influenced by the location of the accretion spot 
on the surface of the white dwarf, which is established in \ob\ to depend  
on the mass accretion rate. Taking the observed accretion geometry changes
from high to intermediate states, the movement of the 
accretion spot from $\psi = 45$\degr\ to $\psi = 35$\degr\
\citep{Schwope01}, we estimate the 
possible, additional shift of the $(O-C)$-values to be only 2 sec,
being thus small compared to the observed deviations.

The observed variations in the $(O-C)$-diagram must either be 
attributed to a secular period variation due to angular momentum 
loss or cyclic variations or a combination thereof. 
The situation remains so far ambiguous
with both a quadratic and sinusoidal ephemeris being equally possible 
and providing a similarly bad or good
description of the data. A combined quadratic and sinusoidal ephemeris 
results in a significantly better fit. 
If true, the observed period decrease $(\dot{P}_{\rm orb} = -7..-11 
\times 10^{-12})$s~s$^-1$ would be large, implying an angular momentum loss 
of $\dot{J} = 3.7\times 10^{35}$erg 
neglecting the effects of mass loss. This is in the range of 
standard magnetic braking \citep{Rappaport83}, but a factor of
30 larger than expected by gravitational radiation alone. Such a strong 
angular momentum loss for a system below the period gap would clearly 
violate the standard evolutionary scenario of disrupted magnetic braking.
Notably, a period decrease at a similar level was recently detected in three 
other short period binaries: DP Leo \citep{Schwope02a}, 
NN Ser \citep{Brinkworth06} and OY Car \citep{Greenhill06}. Future observations
have to show if these variations are indeed secular or  
if their periodic nature is masked by the yet insufficient baseline.

\citet{Borges08} have compiled a selection of 14 eclipsing CVs 
for which a well sampled $O-C$-diagram has been established. All these 
systems show variations of the period, which in the long-run mostly appear
to be cyclic masking any remaining secular trend. 
The ($O-C$)-diagram of \ob\ also reveals clear indication for a cyclic 
change, irrespective of whether a strong secular period evolution is assumed 
or not. The measured modulation periods (13 or 6.9 yrs) and the amplitudes  
imply a fractional period change $\Delta P/P_{\rm
mod}$ of $3-7\times 10^{-7}$, which is comparable to the values of 
the four CVs of \citet{Borges08} that lie below the period gap.
The preferred mechanism invoked to explain these 
cyclic variations in CVs and other long-period close binaries 
(Algol, RS CVn and W UMa stars) are period modulations due to shape changes 
induced by a variable magnetic field \citep{Applegate92}.

A recent recalculation of Applegate's mechanism for the short-period
pre-CVs NN Ser \citep{Brinkworth06} shows that the energy budget of the
low-mass secondary is one order of magnitude lower than the energy
needed to drive the observed period variations. The new calculations
take into account a variable shell mass and the contribution of the
quadrupole moment of the core. Following the calculations of
\citet{Brinkworth06} we tested if the
energy budget of the secondary star in \ob\ allows for the Applegate mechanism at all.
Using M$_2$ = 0.2 \msun, R$_2$ = 0.22 \rsun\ and M$_{\rm{WD}}$ = 0.88 \msun 
(Vogel et al., in prep.) the change in period implied by the sinusoidal
terms of the ephemerides result
in a required energy to drive the Applegate mechanism which exceeds the total
radiant energy of the secondary in the case of the quadratic + sinusoidal ephemeris
by a factor of 3.5 and still requires 80 per cent at minimum in the case of the sinusoidal
ephemeris. Neglecting any angular momentum loss the quadratic ephemeris also
requires more than half $(\sim 70\%)$ of the total energy output of the star. 
We therefore claim that the Applegate mechanism cannot account for the
possible sinusoidal modulation of the orbital period.

As an alternative origin the cyclic period changes could also be caused 
by an unseen third body in which case
the variation should be strictly periodic. Taking the elements 
of the quadratic plus sinusoidal 
ephemeris we compute the minimum mass of such a body assuming that 
its inclination is 90\degr . The observed mass function \citep{Borkovits96} 
is $1.1\times10^{-7}$\Msun\ implying a minimum mass of the third body of
0.0047\Msun\  (5 Jupiter masses) for an assumed total mass of the 
system of 1.0~\Msun .
On the otherhand, a third body more massive than an object at the 
hydrogen-burning limit would require an inclination $<3$\degr\ and 
is therefore not very likely.

\begin{acknowledgements}
RS and JV are supported by the Deut\-sches Zentrum f\"ur
Luft- und Raumfahrt (DLR) GmbH under contracts No. FKZ  
\mbox{50 OR 0206} and \mbox{50 OR 0404}.
ULTRACAM is supported by STFC grant PP/D002370/1.
\end{acknowledgements}
\bibliographystyle{aa}
\bibliography{aamnemonic,myrefs}

\begin{thebibliography}{32}
\expandafter\ifx\csname natexlab\endcsname\relax\def\natexlab#1{#1}\fi

\bibitem[{{Applegate}(1992)}]{Applegate92}
{Applegate}, J.~H. 1992, \apj, 385, 621

\bibitem[{{Beuermann} {et~al.}(2008){Beuermann}, {El Kholy}, \&
  {Reinsch}}]{Beuermann08}
{Beuermann}, K., {El Kholy}, E., \& {Reinsch}, K. 2008, \aap, 481, 771

\bibitem[{{Borges} {et~al.}(2008){Borges}, {Baptista}, {Papadimitriou}, \&
  {Giannakis}}]{Borges08}
{Borges}, B.~W., {Baptista}, R., {Papadimitriou}, C., \& {Giannakis}, O. 2008,
  \aap, 480, 481

\bibitem[{{Borkovits} \& {Hegedues}(1996)}]{Borkovits96}
{Borkovits}, T. \& {Hegedues}, T. 1996, \aaps, 120, 63

\bibitem[{{Bridge} {et~al.}(2002){Bridge}, {Cropper}, {Ramsay}, {Perryman}, {de
  Bruijne}, {Favata}, {Peacock}, {Rando}, \& {Reynolds}}]{Bridge02}
{Bridge}, C.~M., {Cropper}, M., {Ramsay}, G., {et~al.} 2002, \mnras, 336, 1129

\bibitem[{{Brinkworth} {et~al.}(2006){Brinkworth}, {Marsh}, {Dhillon}, \&
  {Knigge}}]{Brinkworth06}
{Brinkworth}, C.~S., {Marsh}, T.~R., {Dhillon}, V.~S., \& {Knigge}, C. 2006,
  \mnras, 365, 287

\bibitem[{{Dhillon} {et~al.}(2007){Dhillon}, {Marsh}, {Stevenson}, {Atkinson},
  {Kerry}, {Peacocke}, {Vick}, {Beard}, {Ives}, {Lunney}, {McLay}, {Tierney},
  {Kelly}, {Littlefair}, {Nicholson}, {Pashley}, {Harlaftis}, \&
  {O'Brien}}]{Dhillon07}
{Dhillon}, V.~S., {Marsh}, T.~R., {Stevenson}, M.~J., {et~al.} 2007, \mnras,
  378, 825

\bibitem[{{Done} {et~al.}(1992){Done}, {Mulchaey}, {Mushotzky}, \&
  {Arnaud}}]{Done92}
{Done}, C., {Mulchaey}, J.~S., {Mushotzky}, R.~F., \& {Arnaud}, K.~A. 1992,
  \apj, 395, 275

\bibitem[{{Frank} {et~al.}(1988){Frank}, {King}, \& {Lasota}}]{Frank88}
{Frank}, J., {King}, A.~R., \& {Lasota}, J.-P. 1988, \aap, 193, 113

\bibitem[{{G{\"a}nsicke} {et~al.}(2006){G{\"a}nsicke}, {Long}, {Barstow}, \&
  {Hubeny}}]{Gaensicke06}
{G{\"a}nsicke}, B.~T., {Long}, K.~S., {Barstow}, M.~A., \& {Hubeny}, I. 2006,
  \apj, 639, 1039

\bibitem[{{Greenhill} {et~al.}(2006){Greenhill}, {Hill}, {Dieters}, {Fienberg},
  {Howlett}, {Meijers}, {Munro}, \& {Senkbeil}}]{Greenhill06}
{Greenhill}, J.~G., {Hill}, K.~M., {Dieters}, S., {et~al.} 2006, \mnras, 372,
  1129

\bibitem[{{Kanbach} {et~al.}(2003){Kanbach}, {Kellner}, {Schrey}, {Steinle},
  {Straubmeier}, \& {Spruit}}]{Kanbach03}
{Kanbach}, G., {Kellner}, S., {Schrey}, F.~Z., {et~al.} 2003, in Presented at
  the SPIE Conference, Vol. 4841, SPIE Conference Series, ed. M.~{Iye} \&
  A.~F.~M. {Moorwood}, 82--93

\bibitem[{{King} \& {Lasota}(1979)}]{King79}
{King}, A.~R. \& {Lasota}, J.~P. 1979, \mnras, 188, 653

\bibitem[{{Lamb} \& {Masters}(1979)}]{Lamb79}
{Lamb}, D.~Q. \& {Masters}, A.~R. 1979, \apjl, 234, L117

\bibitem[{{Leggett} {et~al.}(1996){Leggett}, {Allard}, {Berriman}, {Dahn}, \&
  {Hauschildt}}]{Leggett96}
{Leggett}, S.~K., {Allard}, F., {Berriman}, G., {Dahn}, C.~C., \& {Hauschildt},
  P.~H. 1996, \apjs, 104, 117

\bibitem[{{Magdziarz} \& {Zdziarski}(1995)}]{Magdziarz95}
{Magdziarz}, P. \& {Zdziarski}, A.~A. 1995, \mnras, 273, 837

\bibitem[{{Pizzolato} {et~al.}(2003){Pizzolato}, {Maggio}, {Micela},
  {Sciortino}, \& {Ventura}}]{Pizzolato03}
{Pizzolato}, N., {Maggio}, A., {Micela}, G., {Sciortino}, S., \& {Ventura}, P.
  2003, \aap, 397, 147

\bibitem[{{Pringle}(1975)}]{Pringle75}
{Pringle}, J.~E. 1975, \mnras, 170, 633

\bibitem[{{Ramsay} \& {Cropper}(2003)}]{Ramsay03}
{Ramsay}, G. \& {Cropper}, M. 2003, \mnras, 338, 219

\bibitem[{{Ramsay} \& {Cropper}(2004)}]{Ramsay04}
---. 2004, \mnras, 347, 497

\bibitem[{{Ramsay} {et~al.}(2004){Ramsay}, {Cropper}, {Mason}, {C{\'o}rdova},
  \& {Priedhorsky}}]{Ramsay04c}
{Ramsay}, G., {Cropper}, M., {Mason}, K.~O., {C{\'o}rdova}, F.~A., \&
  {Priedhorsky}, W. 2004, \mnras, 347, 95

\bibitem[{{Rappaport} {et~al.}(1983){Rappaport}, {Verbunt}, \&
  {Joss}}]{Rappaport83}
{Rappaport}, S., {Verbunt}, F., \& {Joss}, P.~C. 1983, \apj, 275, 713

\bibitem[{{Sanz-Forcada} {et~al.}(2003){Sanz-Forcada}, {Brickhouse}, \&
  {Dupree}}]{Sanz-Forcada03}
{Sanz-Forcada}, J., {Brickhouse}, N.~S., \& {Dupree}, A.~K. 2003, \apjs, 145,
  147

\bibitem[{{Schwope} {et~al.}(2004){Schwope}, {Hambaryan}, {Staude}, {Schwarz},
  {Kanbach}, {Steinle}, {Schrey}, {Marsh}, {Dhillon}, {Osborne}, {Wheatley}, \&
  {Potter}}]{Schwope04}
{Schwope}, A., {Hambaryan}, V., {Staude}, A., {et~al.} 2004, in ASP Conf. Ser.
  315: IAU Colloq. 190: Magnetic Cataclysmic Variables, 92--+

\bibitem[{{Schwope} {et~al.}(2002){Schwope}, {Hambaryan}, {Schwarz}, {Kanbach},
  \& {G{\"a}nsicke}}]{Schwope02a}
{Schwope}, A.~D., {Hambaryan}, V., {Schwarz}, R., {Kanbach}, G., \&
  {G{\"a}nsicke}, B.~T. 2002, \aap, 392, 541

\bibitem[{{Schwope} {et~al.}(1997){Schwope}, {Mantel}, \& {Horne}}]{Schwope97}
{Schwope}, A.~D., {Mantel}, K.-H., \& {Horne}, K. 1997, \aap, 319, 894

\bibitem[{{Schwope} {et~al.}(2001){Schwope}, {Schwarz}, {Sirk}, \&
  {Howell}}]{Schwope01}
{Schwope}, A.~D., {Schwarz}, R., {Sirk}, M., \& {Howell}, S.~B. 2001, \aap,
  375, 419

\bibitem[{{Schwope} {et~al.}(1993){Schwope}, {Thomas}, \&
  {Beuermann}}]{Schwope93}
{Schwope}, A.~D., {Thomas}, H.~C., \& {Beuermann}, K. 1993, \aap, 271, L25

\bibitem[{{Vogel} {et~al.}(2008){Vogel}, {Schwope}, {Schwarz}, {Kanbach},
  {Dhillon}, \& {Marsh}}]{Vogel08a}
{Vogel}, J., {Schwope}, A., {Schwarz}, R., {et~al.} 2008, in American Institute
  of Physics Conference Series, Vol. 984, High Time Resolution Astrophysics:
  The Universe at Sub-Second Timescales, ed. D.~{Phelan}, O.~{Ryan}, \&
  A.~{Shearer}, 264--267

\bibitem[{{Vogel} {et~al.}(2007){Vogel}, {Schwope}, \&
  {G{\"a}nsicke}}]{Vogel07}
{Vogel}, J., {Schwope}, A.~D., \& {G{\"a}nsicke}, B.~T. 2007, \aap, 464, 647

\bibitem[{{Warner}(1995)}]{Warner95}
{Warner}, B. 1995, {Cataclysmic variable stars} (Cambridge Astrophysics Series,
  Cambridge, New York: Cambridge University Press, 1995)

\bibitem[{{Woelk} \& {Beuermann}(1992)}]{Woelk92}
{Woelk}, U. \& {Beuermann}, K. 1992, \aap, 256, 498

\end{thebibliography}

\addtocounter{table}{1}
\longtab{2}{
\begin{longtable}{rrll}
\caption{\label{t:timings}
Eclipse egress times of \ob. BJED is the barycentrically corrected
ephemeris time of the eclipse egress}\\
\hline\hline
Cycle & BJED      & $\Delta$BJED & Instrument\\[1ex]
      & $+2440000$ & \\[1ex]
\hline
\endfirsthead
\caption{continued.}\\
\hline\hline
Cycle & BJED      & $\Delta$BJED & Instrument\\[1ex]
      & $+2440000$ & \\[1ex]
\hline
\endhead
\hline
\endfoot
1319 &  9217.4361120 & 0.0000115 & MCCP+CA2.2   \\        
1320 &  9217.5229220 & 0.0000115 & MCCP+CA2.2   \\        
1321 &  9217.6097490 & 0.0000115 & MCCP+CA2.2   \\        
1322 &  9217.6966010 & 0.0000231 & ESO1m  \\        
1333 &  9218.6516100 & 0.0000231 & ESO1m  \\        
1334 &  9218.7384390 & 0.0000231 & ESO1m  \\        
1367 &  9221.603501 & 0.0000231  & ESO1m  \\        
1368 &  9221.690319 & 0.0000231  & ESO1m  \\        
1369 &  9221.777148 & 0.0000231  & ESO1m  \\        
14087 & 10325.959254 & 0.0000277 & HST    \\        
14088 & 10326.046074 & 0.0000277 & HST    \\        
14236 & 10338.895480 & 0.0000277 & HST    \\        
13620 & 10285.414104 & 0.0001157 & AIP    \\        
13621 & 10285.500878 & 0.0001157 & AIP    \\        
13632 & 10286.455890 & 0.0001157 & AIP    \\        
14115 & 10328.390324 & 0.0001157 & AIP    \\        
14116 & 10328.477162 & 0.0001157 & AIP    \\        
14138 & 10330.387041 & 0.0001157 & AIP    \\        
14139 & 10330.473866 & 0.0001157 & AIP    \\        
22788 & 11081.383711 & 0.0001157 & AIP    \\        
27394 & 11481.278636 & 0.0000231 & AIP    \\        
29955 & 11703.625705 & 0.0000810 & AIP    \\        
29966 & 11704.580704 & 0.0000810 & AIP    \\        
35043 & 12145.367925 & 0.0000463 & AIP    \\        
29955 & 11703.625705 & 0.0000925 & CA123  \\        
29966 & 11704.580704 & 0.0000925 & CA123  \\        
31312 & 11821.441021 & 0.0000115 & STJ$^1$    \\        
31313 & 11821.527841 & 0.0000115 & STJ$^1$    \\        
25892 & 11350.874322 & 0.0000115 & Optima \\  
25926 & 11353.826219 & 0.0000115 & Optima \\  
25938 & 11354.868078 & 0.0000115 & Optima \\  
30276 & 11731.495069 & 0.0000115 & Optima \\  
30277 & 11731.581892 & 0.0000115 & Optima \\  
35376 & 12174.279097 & 0.0000115 & Optima \\  
35377 & 12174.365920 & 0.0000115 & Optima \\  
38109 & 12411.559191 & 0.0000115 & Optima \\  
39731 & 12552.381844 & 0.0000115 & Optima \\  
39742 & 12553.336841 & 0.0000115 & Optima \\  
42441 & 12787.665041 & 0.0000115 & Optima \\  
42463 & 12789.575075 & 0.0000115 & Optima \\  
42464 & 12789.661918 & 0.0000115 & Optima \\  
47253 & 13205.444710 & 0.0000173 & Optima \\ 
47254 & 13205.531527 & 0.0000173 & Optima \\ 
47300 & 13209.525279 & 0.0000173 & Optima \\ 
47335 & 13212.563996 & 0.0000173 & Optima \\ 
48265 & 13293.306950 & 0.0000173 & Optima \\ 
48288 & 13295.303800 & 0.0000173 & Optima \\ 
48299 & 13296.258845 & 0.0000173 & Optima \\ 
48334 & 13299.297557 & 0.0000173 & Optima \\ 
51032 & 13533.539015 & 0.0000173 & Optima \\ 
55466 & 13918.500708 & 0.0000173 & Optima \\ 
55546 & 13925.446340 & 0.0000173 & Optima \\ 
60096 & 14320.479250 & 0.0000173 & Optima \\ 
60097 & 14320.566050 & 0.0000173 & Optima \\ 
38105 & 12411.211871 & 0.0000578 & XMM OM \\ 
42352 & 12779.938034 & 0.0000578 & XMM OM \\ 
38107 & 12411.385572 & 0.0000231 & XMM MOS1      \\ 
38107 & 12411.385578 & 0.0000231 & XMM MOS2      \\ 
38108 & 12411.472392 & 0.0000231 & XMM PN      \\ 
38133 & 12413.642851 & 0.0000115 & ULTRACAM  \\  
38145 & 12414.684740 & 0.0000115 & ULTRACAM   \\  
42395 & 12783.671299 & 0.0000115 & ULTRACAM   \\  
50702 & 13504.888294 & 0.0000056 & ULTRACAM \\ 
50713 & 13505.843317 & 0.0000056 & ULTRACAM \\ 
50714 & 13505.930139 & 0.0000056 & ULTRACAM \\ 
50724 & 13506.798342 & 0.0000056 & ULTRACAM \\ 
50725 & 13506.885162 & 0.0000056 & ULTRACAM \\ 
50737 & 13507.927008 & 0.0000056 & ULTRACAM \\ 
59524 & 14270.817989 & 0.0000157 & ULTRACAM \\ 
59525 & 14270.904812 & 0.0000157 & ULTRACAM \\ 
59558 & 14273.769884 & 0.0000157 & ULTRACAM \\ 
59559 & 14273.856704 & 0.0000157 & ULTRACAM \\ 
\noalign{\smallskip} \hline \noalign{\smallskip}
   \noindent\noindent\small $^{(1)}$ From \citet{Bridge02}\\
            
\end{longtable}
}
\end{document}